\documentclass{ecai}
\usepackage{times}
\usepackage{graphicx}
\usepackage{latexsym}

\begin{document}

\title{\textsc{KnowledgeCheckR}: Intelligent Techniques for Counteracting Forgetting}

\author{Martin Stettinger\institute{SelectionArts, email: m.stettinger@selectionarts.com} \and Trang Tran\institute{Graz University of Technology, email: \{ttran, afelfernig, rsamer, muatas\}@ist.tugraz.at} \and Ingo Pribik\institute{Flex Austria, email: ingo.pribik@flex.com} \and Gerhard Leitner\institute{University of Klagenfurt, email: Gerhard.Leitner@aau.at}\\  \and Alexander Felfernig$^2$ \and Ralph Samer$^2$  \and Müslüm Atas$^2$ \and Manfred Wundara\institute{Information Technology Department, Villach, Austria, email: manfred.wundara@villach.at}}

\maketitle
\bibliographystyle{ecai}

\begin{abstract}
Existing e-learning environments primarily focus on the aspect of providing intuitive learning contents and to recommend learning units in a personalized fashion. The major focus of the \textsc{KnowledgeCheckR} environment is to take into account forgetting processes which immediately start after a learning unit has been completed. In this context, techniques are needed that are able to predict which learning units are the most relevant ones to be repeated in future learning sessions. In this paper, we provide an overview of the recommendation approaches integrated in \textsc{KnowledgeCheckR}. Examples thereof are \emph{utility-based recommendation} that helps to identify learning contents to be repeated in the future, \emph{collaborative filtering} approaches that help to implement session-based recommendation, and \emph{content-based recommendation} that supports intelligent question answering. In order to show the applicability of the presented techniques, we provide an overview of the results of empirical studies that have been conducted in real-world scenarios.
\end{abstract}

\section{Introduction}

The concept of \emph{inverted learning} is gaining momentum in different types of educational settings \cite{Lakmal2015}. The one-way delivery of information is replaced by face-to-face interaction with work in small groups and a clarification focus. The major focus of the \textsc{KnowledgeCheckR} environment\footnote{A responsive HTML-5 application: www.knowledgecheckr.com} is to provide intelligent techniques that support inverted learning scenarios. On the basis of recommendation functionalities, the system is able to propose learning content and questions that enable students (learners) to better focus on topics where they need to catch up. Also, teachers have better insights into the performance of students and thus can immediately adapt their focus in onsite teaching sessions.  \textsc{KnowledgeCheckR} is based on recommendation approaches that support learning-related tasks such as \emph{scheduling repetition cycles}, \emph{recommending questions and knowledge units}, and \emph{supporting Q\&A scenarios}. In this paper, we provide an overview of \textsc{KnowledgeCheckR} recommendation approaches and report the results of empirical studies that show in which way recommenders can improve the quality of learning.

A major focus of \textsc{KnowledgeCheckR} is the provision of techniques that help counteracting forgetting \cite{Kang2016}. Our motivation to develop such techniques is based on an empirical study conducted with \emph{N=70 companies} in Austria from diverse domains such as financial services, software development, production, transport, telecommunications, and higher education. Participants of the study ranged from lower, medium, up to higher management. On an average, the study participants reported to spend around \emph{5 hours per week} to answer questions of colleagues although these colleagues should already be able the answer the questions, since they visited topic-related educational programs. The major related knowledge categories are summarized in Table \ref{table:knowledgecategories}. 

We experienced similar results in university contexts where, for example, PhD project relevant knowledge has to be "manually" transferred a couple of times to assure that the knowledge is available when needed. Examples thereof are issues such as how to write papers, what are the correct formulations to explain an example, and how to perform a logical proof. In our study, 91.43\% of the participants agreed that mechanisms that help counteracting the  forgetting of company-relevant knowledge are extremely important and 97.24\% mention that a personalized knowledge transfer for counteracting forgetting is important for the company. These were major motivations that lead to the development of \textsc{KnowledgeCheckR}. 

\begin{table}
\centering
\begin{tabular}{|l|c|c|c|}
\hline
category & example & support 
\tabularnewline
\hline
products 	& what are the new product features? & 43.48\%
\tabularnewline
\hline
norms	& how to communicate with colleagues? & 33.33\%
\tabularnewline
\hline
laws	& general data protection regulation & 36.23\%
\tabularnewline
\hline
internal processes	& how to behave during an evacuation? & 34.78\%
\tabularnewline
\hline
business processes	& production, customer complaints & 65.22\%
\tabularnewline
\hline
\end{tabular} \vspace{-0.25cm}
\caption{Major knowledge categories which could profit from techniques for counteracting forgetting.}
\label{table:knowledgecategories}
\end{table}

\vspace{-0.5cm}

The major contributions of this paper are the following: we show how recommendation technologies can be applied to (1) counteract forgetting processes, (2) recommend relevant learning contents, and (3) support Q\&A scenarios in an intelligent fashion. Furthermore, we report initial results of empirical studies that show the applicability and business relevance of our approach. An overview of existing industrial and university-level deployments of \textsc{KnowledgeCheckR} is provided in Table \ref{table:deployments}.

\begin{table}
\centering
\begin{tabular}{|l|c|c|c|}
\hline
domain 	& description & \#users 
\tabularnewline
\hline
public administration	& 3 municipalities in Austria &  1.000
\tabularnewline
\hline
higher education	& 5 universities in Austria &  100
\tabularnewline
\hline
hardware production	& Flex Austria &  100
\tabularnewline
\hline
medicine	& 1 hospital in Austria &  200
\tabularnewline
\hline
\end{tabular} \vspace{-0.25cm}
\caption{Existing real-world deployments of \textsc{KnowledgeCheckR}.}
\label{table:deployments}
\end{table}


The remainder of this paper is organized as follows. In Section \ref{technologies}, we provide an overview of \textsc{KnowledgeCheckR} recommendation technologies. Thereafter, in Section \ref{ui} we provide examples of the system user interface and discuss the provided functionalities. In Section \ref{studies}, we summarize the results of user studies that show the applicability and business relevance of the \textsc{KnowledgeCheckR} environment. In Section \ref{work}, we provide an overview of related work. We conclude the paper with a discussion of future work in Section \ref{future}.

\section{Recommendation Technologies}\label{technologies}

\textsc{KnowledgeCheckR} recommendation approaches support different goals which can be summarized as (1) recommending \emph{question sequences} \cite{Quadrana2018} (following the paradigm of test-enhanced learning \cite{Roediger2006}), (2) recommending \emph{questions for repetition purposes} in order to be able to counteract forgetting \cite{Kang2016}, and (3) recommending further \emph{learning units} that might be of interest for the user \cite{Pazzani2007}. In \textsc{KnowledgeCheckR}, such scenarios are supported by recommendation techniques. Questions are recommended (1) for \emph{learning purposes} when users start to engage in a learning process and (2) for \emph{repetition purposes} after initial learning has been completed. The former is supported by \emph{session-based recommendation} that guides a user through a learning process with increasing question complexity, the latter by a \emph{utility-based recommendation approach} that identifies questions with a higher probability of already being forgotten.

\emph{Session-based Recommendation}. In many application scenarios, users of \textsc{KnowledgeCheckR} prefer to stay anonymous especially when using the system the first time. In such scenarios, not much information regarding the knowledge level and domain-specific experiences of a user is available. In \textsc{KnowledgeCheckR}, a session-based recommendation approach \cite{Wang2019} is provided where already completed similar (nearest neighbor) sessions (sessions that show a similar user interaction behavior) are applied to recommend the next questions to the current (anonymous) user. In \textsc{KnowledgeCheckR}, the session-based approach is based on collaborative filtering \cite{Ekstrand2010}. A simplified example of the approach is depicted in Table \ref{tab:sessionbasedrec}, where sequences of questions answered by other users are stored in a log. The underlying idea is to find sequences (rankings of questions) which are easy to complete where a question ranking is determined on the basis of the question selection behavior of one or a set of nearest neighbors. This goal can be achieved with collaborative filtering, since (implicit) dependencies between questions (e.g., question $x$ is a precondition of  $y$) can be taken into account by learning from similar sessions.

In the example depicted in Table \ref{tab:sessionbasedrec}, the user in the current session $s_c$ has already successfully answered the questions \{$q_1, q_2, q_3$\} in the order [$q_3, q_2, q_1$] but did not answer the questions \{$q_4, q_5$\}. In this context,  \textsc{KnowledgeCheckR} tries to figure out the ordering (ranking) in which the unanswered questions should be presented to the user. Following a session-based recommendation approach, the system identifies the \emph{n-nearest neighbors} and derives a question sequence that might be of relevance for the user. 

In \textsc{KnowledgeCheckR}, the most similar sessions are used to predict questions of potential relevance to the current (anonymous) user. First, the similarity between the session $s_c$ (the current session) and a session $s_a$ can be determined on the basis of Formula \ref{eq:cf} where $correct(q_i,s_x)$ indicates whether question $q_i$ has been correctly answered in session $s_x$ and $Q$ denotes the complete set of questions in a specific \textsc{KnowledgeCheckR} application. In our example, $sim(s_c, s_1) = 1.0$  since there is a complete overlap in the terms of correct questions already answered in $s_c$.

\begin{equation} \label{eq:cf}
    sim(s_a, s_c) = \frac{|q_i \in Q: correct(q_i,s_a) \land correct(q_i,s_c)|}{|q_i \in Q|}
\end{equation}

Formula \ref{eq:cfeval} helps to determine an overall evaluation of question $q$ in the context of the current session $s_c$ with regard to its relevance for the user. In this context, $r(q,s_i)$ denotes the ranking of question $q$ in the session $s_i$ where $SNN$ denotes the n-nearest neighbor sessions of session $s_c$ (the current session). Assuming $SNN = \{s_1\}$ (for the purpose of our example, we follow a 1-nearest neighbor approach), the overall evaluations of the two up-to-now unanswered questions would be $eval(q_4,s_c)=5$ and $eval(q_5,s_c)=4$ where $r(q_4,s_1) = 5$ and $r(q_5,s_1) = 4$.

\begin{equation} \label{eq:cfeval}
    eval(q, s_c) = \frac{\Sigma_{s_i \in SNN (s_i \neq s_c)} r(q, s_i) \times sim(s_i,s_c)}{|SNN|}
\end{equation}

Finally, Formula \ref{eq:cfpred} helps to determine a prediction for the ranking of question $q$ in the context of session $s_c$. The rank of the last question answered within the scope of session $s_c$ is 3, i.e., $currentqrank(s_c)=3$. Furthermore, $rank(eval(q_5,s_c))=1$ and $rank(eval(q_4,s_c))=2$. As a consequence, $pred(q_4,s_c) = 3+2$ and  $pred(q_5,s_c) = 3+1$ which results in the recommendation of $q_5$ as the next question to be posed to the user in $s_c$. If a user provides a wrong answer within the scope of a learning session, the corresponding question is dropped from the list of recommended questions for a specific time period (the default setting is 20 minutes). 

\begin{equation} \label{eq:cfpred}
    pred(q, s_c) = currentqrank(s_c) + rank(eval(q,s_c))
\end{equation}

\begin{table}
\centering{}\begin{tabular}{|c|c|c|c|c|c|} 
\hline 
session     & $q_1$ & $q_2$ & $q_3$ & $q_4$ & $q_5$  	        \tabularnewline
\hline
\hline
$s_1$ 		&  3	& 2     &  1    & 5     & 4 		       \tabularnewline
\hline
$s_2$ 		&  2	& 1     &  3    & ?     & ? 		       \tabularnewline
\hline
$s_3$ 		&  1	& 2     &  4   & 5     & 3 		       \tabularnewline
\hline
$s_4$ 		&  3	& 2     &  4    & 1   & 5 		       \tabularnewline
\hline
$s_c$ 		&  3	& 2     &  1    & ?     & ? 		       \tabularnewline
\hline
\end{tabular}  
\caption{A simple interaction log in session-based recommendation. The table entries represent session-specific orderings of posed questions $q_i$, question marks represent still unknown orderings.} \label{tab:sessionbasedrec}
\end{table}

In our simplified example, the nearest neighbor of session $s_c$ is session $s_1$, since the order in which questions have been answered by the users are the same. The ordering for the next questions that will be recommended by the system is [$q_5, q_4$]. This session-based collaborative approach is used to support the ramp-up in situations where users interact with the learning application maybe the first time and prefer to apply \textsc{KnowledgeCheckR} in an anonymized fashion. After the knowledge level of individual users becomes more transparent, \textsc{KnowledgeCheckR} is able to switch to a \emph{utility-based recommendation approach} where the utility of individual questions is estimated depending on time intervals since questions have been answered correctly the last time.

\emph{Utility-based Recommendation}. In \textsc{KnowledgeCheckR}, utility-based recommendation \cite{Huang2011} is applied to implement functions that help \emph{counteracting forgetting}. The underlying idea is that answering the same question repeatedly within specific time intervals helps to consolidate the learning material \cite{Kang2016}. Furthermore, utility-based recommendation determines a ranking where the most relevant repetitions are presented first. Users under time pressure can thus focus on the most relevant topics.

The functions used to determine questions of relevance are the following. The \emph{relevance} (\emph{rel}) of a question $q$ for a user $u$ is defined as the complement of the share of correct answers compared to the number of answers to question $q$. This factor is weighted by a time aspect, i.e., the more days already passed since the last time the question $q$ has been answered by user $u$  ($\#dayssince(q,u)$), the higher the corresponding relevance since the probability becomes higher that a user is not able to answer the question correctly.  In this context, $\#daystoforget(q)$ represents the assumption that after $x$ days the correct answer will have been forgotten. This value can be pre-specified when defining a question or approximated based on historical data. Equation \ref{eq:rel} represents a basic way of ranking questions.

\begin{equation}\label{eq:rel}
    rel(q,u) = (1 - \frac{\#correctans(q,u)}{\#totalans(q,u)}) \times \frac{\#dayssince(q,u)}{\#daystoforget(q)}
\end{equation}

Equation \ref{eq:relprime} is an extension of Equation \ref{eq:rel} which additionally takes into account the aspects of \emph{importance} and \emph{complexity} of a question (see the Equations \ref{eq:easiness} and \ref{eq:importance}). This means that the higher the \emph{importance} of a question and the lower the \emph{complexity} of a question, the higher the probability that this question will be recommended to the current user. In this context, \emph{complexity} and \emph{importance} can be estimated on the basis of \emph{data related to the global share of correct answers to a specific question} $q$ and the \emph{average feedback of users regarding the importance level of question} $q$. The higher the importance and the lower the complexity, the higher the relevance of the question for a specific user. The underlying idea is that knowledge about simple contents/questions is the precondition for answering more complex ones.

\begin{equation} \label{eq:relprime}
    rel'(q,u) = rel(q,u) \times \frac{importance(q)}{complexity(q)}
\end{equation}

\begin{equation} \label{eq:easiness}
    complexity(q) = 1-(\frac{\#correctans(q)+1}{\#totalans(q)+1})
\end{equation}

\begin{equation} \label{eq:importance}
    importance(q) = \frac{\Sigma_{i=1}^{\#feedbacks(q)} feedbackval(i)}{\#feedbacks(q)+1}
\end{equation}

\emph{Content-based Recommendation}. In learning apps with a large amount of questions, content-based recommendation \cite{Pazzani2007} is used to support intelligent Question \& Answering (Q\&A) which is an orthogonal way to exploit questions and answers stored in  \textsc{KnowledgeCheckR}. The underlying scenario is, for example, the following: a user of a \textsc{KnowledgeCheckR} learning application on \emph{model-based diagnosis} is currently preparing for the exam related to the course. Just two hours before the exam, the question comes to his/her mind, \emph{which diagnosis approach can guarantee the retrieval of minimal cardinality diagnoses}. Such queries can be entered to the search interface of the system. On the basis of the query string (e.g., \emph{which diagnosis approach does support minimal cardinality?}), a content-based recommender determines the similarity between question features and corresponding query features. The answers to questions most similar to the query are then shown to the current user. In order to determine the similarity between the query $q_c$ posed by the current user and a question $q_i \in Q$, the following basic similarity metric is applied (see Formula \ref{eq:contentbasedsim}) \cite{Felfernigetal2018,Pazzani2007}.

\begin{equation} \label{eq:contentbasedsim}
    sim(q_i,q_c) = \frac{2 \times |features(q_i) \cap features(q_c)|} {|features(q_i)| + |features(q_c)|}
\end{equation}

A simplified example of the content-based recommendation approach in \textsc{KnowledgeCheckR} is given in Table \ref{tab:contentbasedrec}. Note that such recommendation services are provided in individual learning applications but as well on the global level to support situations where users are not completely sure which of the available learning applications could answer their questions. In the following example (Table \ref{tab:contentbasedrec}), we assume that the user poses the query in the context of a specific learning application (\emph{model-based diagnosis}). In our example, the question with the highest similarity to the query $q_c$ is $q_6$ ($sim(q_6,q_c) = \frac{2 \times 3}{9} = 0.67$). Consequently, the answer specified for $q_1$ would be shown as answer for $q_c$. For further details on content-based recommendation approaches we refer to \cite{Pazzani2007}.

\begin{table}
\centering{}\begin{tabular}{|c|c|c|c|c|c|} 
\hline 
question     & question features (and question features)	        \tabularnewline
\hline
\hline
$q_1$ 		&  conflict, algorithm, QuickXPlain 		       \tabularnewline
\hline
$q_2$ 		&  FastDiag, algorithm, time, complexity 		       \tabularnewline
\hline
$q_3$ 		&  FastDiag, algorithm, space, complexity 		       \tabularnewline
\hline
$q_4$ 		&  hitting, set, search, tree, breadth, first, minimal, cardinality 		       \tabularnewline
\hline
$q_5$ 		&  hitting, set, conflict, symmetry 		       \tabularnewline
\hline
$q_6$ 		&  minimal, cardinality, diagnosis, search 	       \tabularnewline
\hline
$query (q_c)$ 	&  diagnosis, approach, support, minimal, cardinality	       \tabularnewline
\hline
\end{tabular}  
\caption{\textsc{KnowledgeCheckR}: a simple example of a content-based recommendation setting.} \label{tab:contentbasedrec}
\end{table}

\vspace{-0.25cm}

\section{\textsc{KnowledgeCheckR} User Interfaces}\label{ui}

\begin{figure}[ht]
\centering
\includegraphics[scale=0.225]{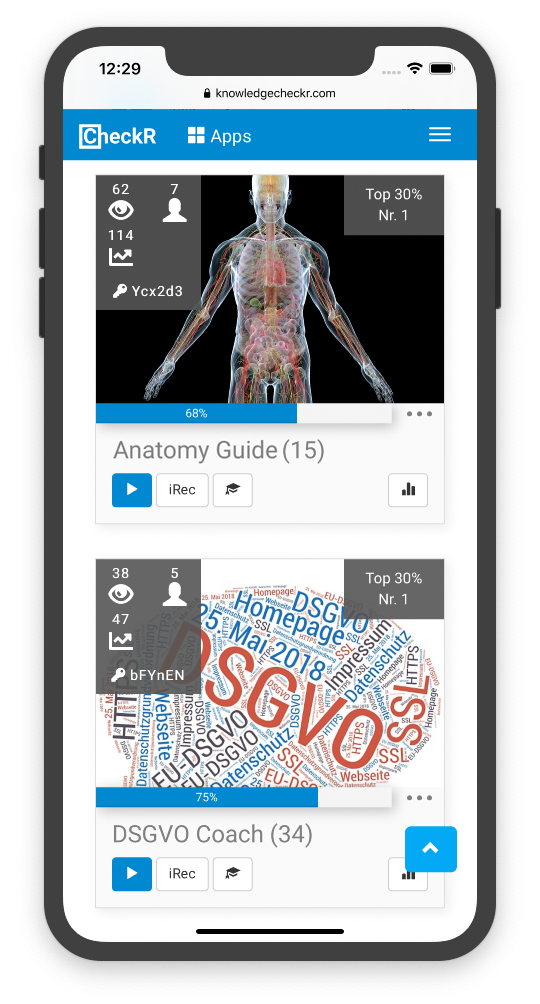}  
\caption{\textsc{KnowledgeCheckR} view on learning apps.} 
\label{fig:apps}
\end{figure}

\begin{figure}[ht]
\centering
\includegraphics[scale=0.225]{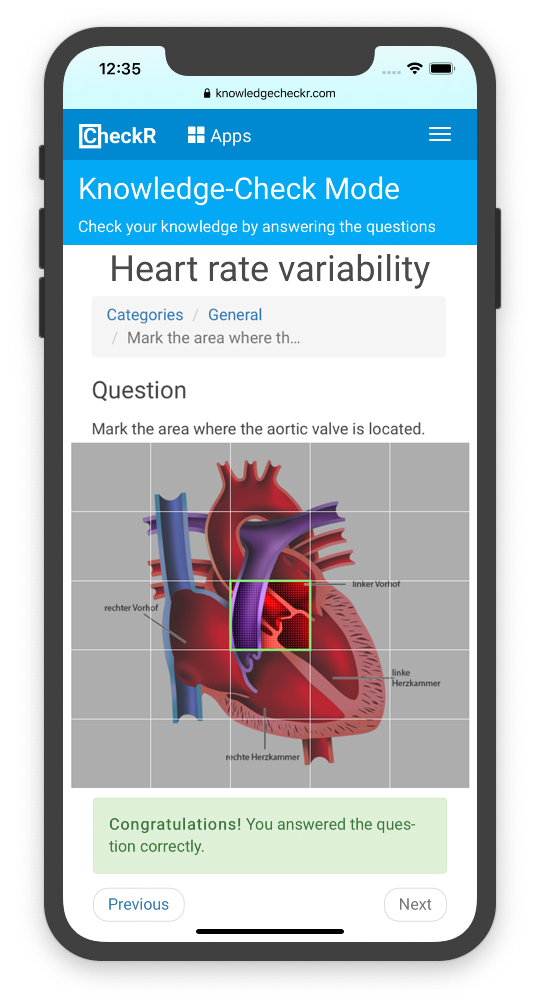}  
\caption{\textsc{KnowledgeCheckR} graphical interaction mode (find the areas in question): an example from the medicine domain where heart parts in question have to be identified.} 
\label{fig:heartinterface}
\end{figure}

\textsc{KnowledgeCheckR} provides the possibility to create individual learning apps (see Figure \ref{fig:apps}) that consist of contents such as movies and slides and a corresponding set of questions that can be used for personal self tests (knowledge checks), learning sessions, exercises, exams, and competitions. The supported question types are \emph{multiple choice}, \emph{sequencing tasks}, \emph{text completion tasks}, and \emph{image analysis tasks} (see also Figure \ref{fig:heartinterface}). For each question type, a corresponding explanation can be defined that is shown (if activated) if a user is not able to provide a correct answer to a question. 

The discussed recommendation approaches (see Section \ref{technologies}) are useful especially in the context of learning sessions where users try to answer questions in order to improve their knowledge level in specific categories. Since questions have different knowledge levels, the recommender system helps a user to focus on questions with a high probability of being answered before being forwarded to more complex ones. Figure \ref{fig:apps} provides an overview of a \textsc{KnowledgeCheckR} list of learning applications -- one example of such an application is the \emph{Anatomy Guide} app which is used in medical domains. Registered users dispose of the additional service of \emph{repetitive recommendations} where questions already posed in the past are posed again in order to achieve the goal of counteracting the \emph{forgetting curve} \cite{Pashler2007}. Each learning app can provide learning content in a personalized fashion and also proposes questions a user should try to answer in the next learning iteration \cite{Roediger2006}. 

As already mentioned, \textsc{KnowledgeCheckR} provides different ways of asking questions. Figure \ref{fig:heartinterface} provides an example of an image recognition task where the task of a user is to answer a medical question by selecting the corresponding image areas which represent the answer. If a question could not be answered correctly, a corresponding explanation can be shown -- in the case of images, an explanation is a visualization of the correct answer areas in the image (including a textual explanation as to why the shown area is the correct one).

Finally, \textsc{KnowledgeCheckR} includes a user interface where the expertise of the community and also of individual users can be analyzed. The interface supports a fine-grained analysis of critical knowledge areas where there is a need to improve the community knowledge or the knowledge of individual users (if the parametrization of system allows this). The analysis section entails a motivational aspect since, for example, the personal comparison with the whole community immediately leads to more system interaction with the goal to be at least as good or even better than the average performance of the community. \textsc{KnowledgeCheckR} also provides mechanisms to directly update participants of a learning application with regard to new contents and questions. This update channel can be configured in terms of the way notifications are explained (see Section \ref{studies}) and the frequency of knowledge updates.

\begin{figure}[ht]
\centering
\includegraphics[scale=0.225]{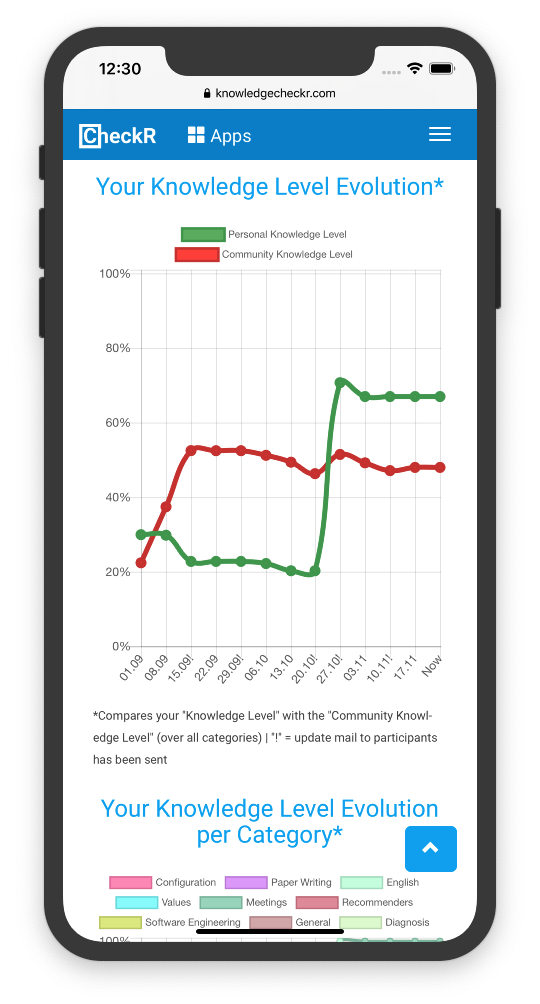}  
\caption{\textsc{KnowledgeCheckR} view on the development of the personal and community knowledge level.} 
\label{fig:knowledgelevel}
\end{figure}

\section{User Studies and Benefits}\label{studies}

As already mentioned, \textsc{KnowledgeCheckR} has already been deployed and is applied in a couple of application scenarios (see Table \ref{table:deployments}). The system is applied for various purposes out of which we will discuss a couple of aspects in the following. 

\emph{Public Administration.} In public administration, the system is applied in e-learning scenarios related to topics such as \emph{safety and security}, \emph{dealing with computers}, \emph{programming best practices}, \emph{requirements engineering best practices}, and \emph{sensitization of citizens}. Whereas the former scenarios focus on knowledge transfer directly to employees of the public administration, the latter one focuses on knowledge transfer between public  administration and citizens. Examples thereof are topics such as \emph{healthy eating behavior}, \emph{environmental protection}, and \emph{first aid}. Especially \textsc{KnowledgeCheckR} \emph{competitions} can be regarded as a question-driven learning channel \cite{Roediger2006} where the questions (and corresponding answers) are a major mean to increase the sensitiveness of citizens with regard to the mentioned topics. Motivation in this context is not self-intrinsic and must be stimulated on the basis of remuneration mechanisms such as prizes provided by companies. 

\begin{figure*}
\centering
\includegraphics[scale=0.55]{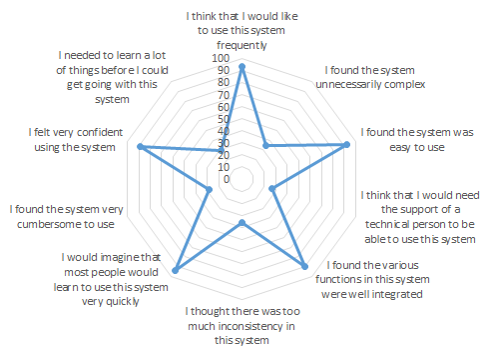}  
\caption{System Usability Scale (SUS) \cite{Bangor2008} evaluation in public administration (evaluation scale 0: \emph{strongly disagree} .. 100: \emph{strongly agree}).} 
\label{fig:sus}
\end{figure*}

We have conducted a \emph{usability study} that focused on public administration end-users of the \textsc{KnowledgeCheckR} environment. The questionnaire was based on the \emph{System Usability Scale} (SUS) \cite{Bangor2008} with N=20 participants providing feedback on the usability of the system in the context of the mentioned administration-internal applications. Overall, the participants of the study provided positive feedback regarding  general usability aspects summarized via SUS questionnaire (see Figure \ref{fig:sus}). Benefits from the application of \textsc{KnowledgeCheckR} in the public administration are (1) reduced efforts for managers to keep their team up-to-date and (2) the avoidance  of cost-intensive mistakes. On an average, time savings due to reduced "update efforts" were reported to be around \emph{2-3 weeks per year}.

\emph{University Courses.} Experiences from the application of \textsc{KnowledgCheckR} in the university context provide similar results. Leaders of research teams experience similar effort reductions (\emph{2-3 weeks per year}) related to the update of their students. An example thereof are systematic updates regarding  basics of their research topics, practices when writing papers, and criteria and strategies for successfully completing their PhD studies. In the context of university courses, \textsc{KnowledgeCheckR} has been applied in computer science teaching. The related learning applications provide an additional means for students to prepare for courses and check their knowledge level in different categories. This is an extremely important feature for students since the system enables them to focus their learning effort on relevant topics in which they have to catch up. 

When used as an additional means to understand course topics and to prepare for an exam, around 10\% of the students use \textsc{KnowledgeCheckR} from the very beginning throughout the whole course. Furthermore, around 80\% of the students primarily use the system directly ahead of an exam with goal to check their knowledge level and to be optimally prepared. Finally, on an average 20\% of the students do not use \textsc{KnowledgeCheckR} at all. Consequently, the usage in scenarios where users are not forced to use the system (contrary to industrial contexts), follows an \emph{80-20 rule}. Similar percentages have been observed in four different courses. 

Finally, we measured the prediction quality of the utility-based approach discussed in Section \ref{technologies}, since in most of the university courses students are used to sign-in (non-anonymous mode) to interact with the system. On an average, the prediction quality of the utility-based recommendation approach in terms of precision \cite{Herlocker2004} is around 0.9 overall all learning apps, i.e., in nearly 90\% of the cases, the system manages to predict the item that will also be chosen by the user. 

In a course with N=350 Computer Science students, we measured potential increases in student output quality that can be explained by the application of \textsc{KnowledgeCheckR}. This evaluation was performed in the context of the course \emph{Object-oriented Analysis and Design} where topics and support team did not change over three years and a significant improvement could be observed in terms of student grades. Compared to previous years, there was a significant reduction of negative grades (evaluation "insufficient") ($6.42$\%) and a significant increase of excellent grades (evaluation "very good") after \textsc{KnowledgeCheckR} has been provided  ($5.5$\%). The previously discussed \emph{80-20 rule} could be confirmed. In this scenario, 350 users answered around $50.000$ questions which means around $143$ answered questions per user.

\emph{User Motivation and Explanations.} User motivation is an important aspect since it is the precondition for a wide-spread application of the system. First, \textsc{KnowledgeChckR} provides functionalities that help to inform participants of learning applications in cases where new contents have been entered into the system or existing ones have been updated. It is important to know that depending on the learning domain, the formulation of related persuasive explanations (arguments) should differ. In \emph{high-involvement learning domains} (users have a high interest in understanding the learning content) such as university courses, persuasive explanations should follow the line of \emph{socialness} (in the line of the persuasion dimensions proposed in Cialdini \cite{Gialdini2014}). An example thereof is the explanation \emph{other users who answered the following questions correctly, managed to pass the exam in 95\% of the cases}. Vice-versa, in \emph{low-involvement learning domains} (users have a low or nearly no interest in understanding the learning content), such as \emph{learning fire protection rules}, instead of socialness, \emph{time-related arguments} seem to be more important. An example of such an argument is the following: \emph{the answering of the following six questions takes only three minutes}.

\section{Related Work}\label{work}

\emph{Recommender Systems.} Recommender systems are used to retrieve items of relevance for the user from a large and potentially complex item assortment \cite{Jannachetal2010}. \emph{Collaborative filtering} \cite{Ekstrand2010} exploits the preferences of so-called nearest neighbors, i.e., users with preferences similar to the current user, and recommends items that have already been consumed (and rated positively) by the nearest neighbors but not by the current user. \emph{Utility-based recommendation} is based on a utility analysis of different items using a utility function \cite{Huang2011}. \emph{Content-based recommender systems} \cite{Pazzani2007} focus on evaluating the similarity between a new item a user did not notice up to now and the user profile derived from previous item consumptions. Finally, \emph{knowledge-based recommender systems} \cite{Burke2000,FelfernigBurke2008} focus on the recommendation of items characterized by attributes where the recommendation knowledge is often described either in terms of constraints or in terms of similarity metrics. 

\emph{Recommender Systems in e-Learning.} Overall, the application of recommender systems in e-learning scenarios primarily focuses on the personalized provision of learning content -- for an overview of recommender systems in e-learning we refer to \cite{Klasnja2015,Krauss2016}. In \textsc{KnowledgeCheckR}, collaborative filtering is applied in the context of providing (session-based) recommendation services to anonymous users, utility-based recommendation is used to implement functionalities to counteract forgetting, and content-based recommendation is included to provide basic Q\&A services. Thus, the application of recommender systems is extended by specifically taking into account requirements of inverted learning processes \cite{Lakmal2015}. 

\emph{Counteracting Forgetting.} Research focused on the analysis of forgetting processes can primarily be found in the psychological literature \cite{Kang2016,Pashler2007,Roediger2006}. For example, Pashler et al. \cite{Pashler2007} analyze possibilities to enhance learning processes and retarding forgetting and point out clear improvements that can be achieved when providing related technologies. In the context of recommender systems, forgetting processes are primarily taken into account in models that represent (long-term) preference shifts \cite{Matuszyk2018}. In this context, knowledge about forgetting processes is exploited to infer preference shifts whereas knowledge about forgetting processes in  \textsc{KnowledgeCheckR} is used to develop strategies that help to counteract forgetting. 

\emph{Explanations and Recommender Systems.} The selection of explanation types implemented in a recommender system strongly depends on the overall goal of the explanation \cite{TintarevMasthoff2011}. Examples of such goals are \emph{increasing the purchase probability of specific items}, \emph{increasing a user's item domain knowledge}, \emph{persuading a user to take specific actions}, and \emph{increasing the trust level of a user}. A major focus of explanation approaches in the \textsc{KnowledgeCheckR} environment is (1) to persuade for system usage and (2) to provide explanations in situations where a user is not able to answer a question correctly. In both cases, the overall impact of these explanations is that users increase their domain-specific knowledge.

\section{Conclusions \& Future Work}\label{future}

\textsc{KnowledgeCheckR} is a learning environment that supports question-enhanced learning processes that enable counteracting forgetting on the basis of recommendation technologies. In this paper, we provided an overview of the algorithmic approaches integrated in \textsc{KnowledgeCheckR} and also discussed example aspects of the user interface provided by the system. In this context, we also reported results from empirical studies conducted on the basis of real-world deployments of \textsc{KnowledgeCheckR}.

Our plans for future work include further extensions of \textsc{KnowledgeCheckR}. First, we plan to integrate automated video segmentation functionalities that help to cut sequences from videos that best help to explain content/question-specific aspects. Second, we will further improve the predictive quality of question recommendations. Third, sentiment learning from chats will be used to estimate more precisely different dimensions such as quality and complexity of a question. Furthermore, we plan to include services such as the optimization of the working load of a user to achieve specific goals. For example, we will provide mechanisms that recommend learning items that have to be "consumed" to achieve specific learning goals such as \emph{successfully passing an exam with minimum effort}. Finally, we plan to analyze in more detail the impact of repetitions on the personal knowledge level evolution.

\ack The work presented in this paper has been conducted within the scope of the \textsc{KnowledgeCheckR} research project at the Graz University of Technology and various related industry cooperations.

\bibliography{ecai}
\end{document}